\documentclass[granma]{svjour}  %%referee
\usepackage{graphics}
\usepackage{color}
\begin{document}
\def\uslide {v_{x,w}/\sqrt{T_{xx,w}}}

 \title{Influence of lateral confinement on granular flows: comparison between shear-driven and gravity-driven flows}

\author{Patrick RICHARD$^1$ and Riccardo ARTONI$^1$ and Alexandre VALANCE$^2$ and Renaud DELANNAY$^2$}

\institute{$^{1}$ MAST-GPEM, Univ Gustave Eiffel, IFSTTAR, F-44344 Bouguenais, France\\
$^2$ Univ Rennes, CNRS, IPR (Institut de Physique de Rennes)–UMR 6251, F-35000 Rennes, France}

\date{Received: date / Revised version: date}
\maketitle
\begin{abstract}
The properties of confined granular flows are studied through discrete numerical simulations.
Two types of flows  with different boundaries are compared:
(i) gravity-driven flows topped with a free surface and over a base where erosion balances accretion (ii) shear-driven flows with a constant pressure applied at their top and a bumpy bottom moving at constant velocity. In both cases we observe shear localization over or/and under a creep zone.
We show that, although the different boundaries induce different flow properties ({e.g.} shear localization of transverse velocity profiles), the two types of flow share common properties like (i) a power law relation between the granular temperature and the shear rate (whose exponent varies from $1$ for dense flows to $2$ for dilute flows) and (ii) a weakening of friction at the sidewalls which gradually decreases with the depth within the flow. 
%This weakening can be explained by the intermittent motion of the grains in creep zone of the flow.
\end{abstract}

\section{Introduction}
A lot of examples of 
confined granular flows can be found both in nature and in industry, from geophysical flows confined by a canyon to grain transport in channels.\\ 
Such types of flows are complex systems because confinement (e.g. top, bottom or sidewalls) may  induce correlations as well as  non-local effects 
%induced by "sidewalls"
that possibly have an influence over long distances~\cite{Jop_JFM_2005}. Also, confined granular flows are likely to develop zones without shear and, consequently, they can experience erosion and accretion~\cite{Taberlet2004b}, which are still the subject of active research~\cite{Farin_JGR_2014,Lefebvre_PRL_2016,Jenkins_PRE_2016,Trinh_PRE_2017}. Therefore they are good systems to test theories dealing with both a solid and a fluid granular phases and how to handle the corresponding phase transition~\cite{Vescovi_SM_2016,Vescovi_GM_2018}.
Also, if one of the ultimate goal of the physics of granular materials is to obtain a full 3D rheological model capturing the behaviour of granular flows, this model has to be fed by boundary conditions at sidewalls (velocity, granular temperature\ldots). Studying confined flows, experimentally and/or numerically, can help to reach this goal by providing the aforementioned conditions.\\
Recently, we have studied steady and fully developed  (SFD) granular flows in two confined geometries: a laterally confined chute flow~\cite{Taberlet2004b,Richard_PRL_2008,Taberlet2008,Holyoake_JFM_2012,Brodu_PRE_2013,Brodu_JFM_2015} and a constant-pressure  confined shear cell for which shear is imposed by a moving bumpy bottom~\cite{Artoni_PRL_2015,Artoni_JFM_2018}.
In the remainder of the paper we will refer to these two types of flows  as gravity-driven flows and shear-driven flows respectively. 
\textcolor{black}{As it will be shown below, each type of flow displays several zones in which the behaviour is specific.
%We will mainly focus on three zones within our systems. First, the shear zone \textit{i.e.} the zone in which the shear is localized. % and the grains flow continuously.
%Second, the creep zone where the grains move in an intermittent way as described in~\cite{Komatsu_PRL_2001,Crassous_JSTAT_2008}. Third,
%as it will be shown below, the shear-driven flow may experience a plug flow. The corresponding zone is refereed as plug zone.
%Note also that, for gravity-driven flows, a gazeous flow might occur atop the shear zone. Yet its study is out of the scope of the present paper.
}\\
%\textcolor{magenta}{The exact boundary between the shear and  creep zone is a matter of debate since creep is not incompatible with shear~\cite{Komatsu_PRL_2001}. This point will be discussed in section~\ref{sec:shear}}.\\
These two geometries have in common the presence of confinement, but they also have important differences like the presence or absence of a free surface and the type of driving force \textcolor{black}{which can be either volumetric --for the former-- of induced by a wall --for the latter--}.\\
In this paper, by using  Discrete Element Method simulations (DEM), we will compare the results obtained in the two aforementioned geometries and discuss the differences and the similarities.\\
The outline of the paper is the following. In Sect.~\ref{sec:geom} we will first describe the two geometries that have been used. Then, in Sect.~\ref{sec:shear}  we will focus on the flow velocity and the shear localization. Section~\ref{sec:T} is devoted to the study of granular temperature, quantity that is very sensitive to non-local effects~\cite{Zhang_PRL_2017}. Sidewall friction and its relation with sliding velocity are discussed in Sect.~\ref{sec:friction}. Finally, we will conclude and discuss some perspectives of the presented work.

\section{Geometries}\label{sec:geom}

In this section we will describe the two configurations used in this work to simulate confined flows of spherical grains: (i) a chute flow confined between sidewalls leading to gravity-driven flows and (ii) a confined shear cell leading to shear-induced flows.  As it will be explained below, in both geometries flow is confined between two sidewalls but they differ by the way flow is driven as well as  by their boundary conditions at the top of flow (free surface and a bumpy bottom submitted to a constant pressure respectively) and at the bottom (bumpy static bottom and moving static bottom respectively). 
%\textcolor{red}{Let us precise here that we have used two different numerical methods for the two geometries \textit{i.e.} soft-sphere molecular dynamics simulations for gravity-driven flows and contact-dynamics for shear-driven flows.
{The latter configuration leads to \textcolor{black}{systems which are dense everywhere} (see Sect.~\ref{sec:shearcell}). This is not the case for gravity-driven flows that are topped by a very dilute region. Also the behaviour of dense flows is weakly dependant on the elastic properties of grains~\cite{Rajchenbach2000}. For this reason, and to save computation time,  the grains involved in shear-driven flows are perfectly inelastic ({i.e.} their restitution coefficient is zero). We also used the Contact-Dynamics method~\cite{Jean_ComputerMethodsApplMechEng_1999} which handles easily purely inelastic grains. In contrast, for gravity-driven flows, low volume fractions are achieved, thus we have used a grain restitution close to that used in experiments~\cite{Taberlet_PRL_2003,Richard2020} and soft-sphere molecular dynamics method. Note however that the effect of the coefficient of restitution is weak and only measurable close to the free surface of the flow in the dilute region.}
It should also be pointed out that the height of the granular system does not play a significant role as long as it is large enough to observe the creeping region. In both geometries, if it is too small, the  system is sheared along its whole depth. 
\subsection{Laterally confined chute flow}
\begin{figure}%[htbp]
\begin{center}
\resizebox{0.85\hsize}{!}{%
  \includegraphics{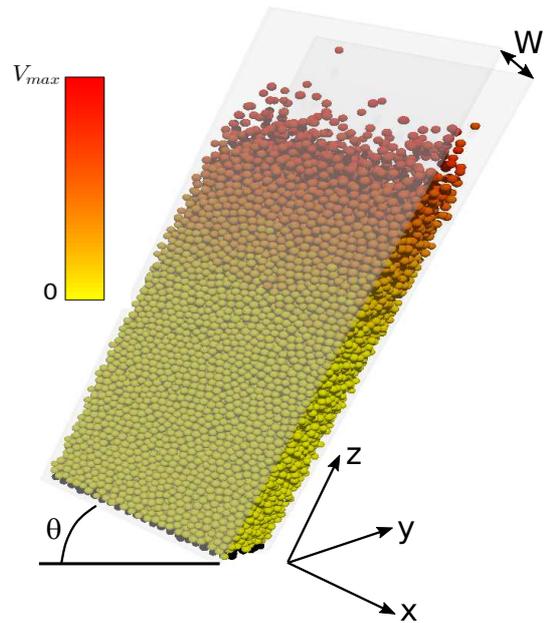}
	}
\caption{Typical 3D snapshot for gravity-driven flows: $W/d=10$ and $N=15,000$. The angle of the flow is $\theta=40^\circ$, the coefficient of
restitution $e_n$ is equal to $0.88$, and the grain-grain and grains-sidewall friction coefficients (respectively $\mu_{pp}$ and $\mu_{pw}$) are set to $0.5$. Flow is directed
down the incline along the $x$-axis. Two sidewalls are parallel to the $(xz)$ plane and confine the system}\label{fig:sketch_SSH}
\end{center}
\end{figure}

The geometry described in the present section aims to model so-called sidewall stabilized heaps (SSH)~\cite{Taberlet_PRL_2003,Taberlet2004b,Richard_PRL_2008,Taberlet2008}
which  are characterized by the
presence of a steep heap %(creep zone) 
beneath a flowing layer. % (shear zone).
It consists in an inclined $3D$ cell   (see Fig.~\ref{fig:sketch_SSH})  similar to those used in~\cite{Taberlet2004b,Richard_PRL_2008,Taberlet2008,Gollin_GM_2017,Berzi_SM_2019,Berzi_JFM_2020,Richard2020}. The angle between the horizontal and the main flow direction ($x-$direction) is called $\theta$.
It has been shown that, in such type of geometries, the angle of the flow is linked to the flow rate as long as there are enough grains in the system to ensure the presence of a creep zone above which a  flow occurs~\cite{Taberlet2008,Richard2020}.
Also such types of flows are influenced by the confinement even at very large widths~\cite{Jop_JFM_2005}.
The size  of the cell in the $x-$ direction is set to $L_X\approx 25d$ with periodic boundary conditions in this direction.
In the $z-$direction ({i.e.} normal to the free surface of the flow) the size of the cell 
is set to large values and thus considered as infinite.
In the $y-$direction, the flow is confined by two flat frictional sidewalls located at positions $y=-W/2$ and $y=W/2$ with $W=10d$. 
The  bottom of the cell is made bumpy by pouring under gravity $\mathbf{g}$ a large number of grains in the cell and by gluing those that are in contact with the plane $z=0$ and removing the others. \\
For confined chute flow simulations, we use soft-sphere molecular dynamics simulations developed internally~\cite{Richard_PRL_2008,Richard_PRE_2012} for which $N=15,000$ grains in contact overlap slightly. The interactions between two grains have both a normal and a tangential component. The normal force, $F_n$, is classically modelled by a spring and a dashpot: $F_n=k_n \delta - \gamma_n \dot \delta$ where $k_n$ and $\gamma_n$ are respectively the stiffness of the spring and the viscosity of the dashpot, $\delta$ the overlap between grains and $\dot\delta$ its derivative with respect to time. The stiffness is set to $5.6 \times 10^6\mbox{ mg/d}$ and we choose the value of $\gamma_n$ such as the normal restitution coefficient is equal to $0.88$~\cite{Richard2020}. Note that the stiffness used is relatively large since a grain located at the bottom of simulation cell topped with a column of grains whose height is similar to that of the flow ({i.e.} between 50 and 100 grain sizes)  has a deformation roughly equal to $10^{-5}d$, {i.e.} much lower than grain size.  
The tangential force is modelled by a spring, $F_t=k_t u_t$, where $k_t=2 k_n/7$. Its deformation $u_t$ ({i.e.} the elastic tangential displacement between grains) is bounded to satisfy Coulomb law $F_t=\mu F_n$, where $\mu$ is the friction coefficient which, in the remainder of the paper, is set to $\mu=\mu_{pp}=0.5$ for a grain-grain contact. 
The walls are treated like spheres of infinite mass and radius. 
The normal restitution coefficient of the grain-wall interaction is the same than that used for the grain-grain interactions. In contrast the value of the friction coefficient between the grains and the walls, $\mu_{pw}$, will be varied to study its effect.
To avoid any structural ordering the diameter of the grains is uniformly distributed between $0.8 \langle d \rangle $ and $1.2 \langle d \rangle$ where $\langle d \rangle$ is the average grain diameter. At the beginning of the simulation the kinetic energy of the system is set to an important value~\cite{Richard2020} such as the SFD state obtained after a transient does not show any sign of the initial structure.

\subsection{Shear Cell}\label{sec:shearcell}
\begin{figure}%[htbp]
\begin{center}
\resizebox{0.85\hsize}{!}{%
  \includegraphics{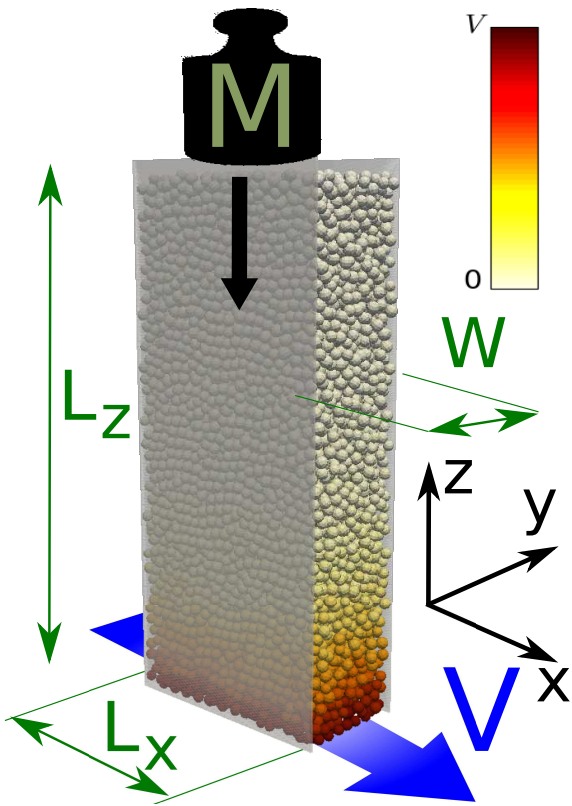}
	}
\caption{Typical 3D snapshot for shear-driven flows.  $N=10,000$ grains flowing in a shear cell made of a bumpy bottom, an bumpy top and two sidewalls (length $L_z$, length $L_x$) separated by a gap $W/d=10$. The shear is ensured by the bumpy bottom which moves at a constant velocity $V$ along the $x-$direction. The top of flow is a bumpy horizontal wall submitted to a force $M\mathbf{g}$}\label{fig:sketch_shear}
\end{center}
\end{figure}

In the shear cell geometry, simulations are performed by
using the LMGC90 open source framework~\cite{Renouf_JCAM_2004} which is based on the contact dynamics method~\cite{Jean_ComputerMethodsApplMechEng_1999}.
The grains are characterized by an infinite stiffness and the forces between grains are determined through an implicit resolution of both the Signorini condition and the Coulomb law at contact.
The flow configuration, sketched in Fig.~\ref{fig:sketch_shear}, is made of a rectangular cuboid
(length $L_x = 20d$, width $W =10d$, and variable height $L_z$) with periodic boundary conditions along the main flow direction ($x-$direction).
Two flat but frictional lateral  sidewalls (normal to
the $y-$direction and located at positions $y=-W/2$ and $y=W/2$ with $W=10d$) and two horizontal  bumpy walls (at the top and bottom of flow) confine the system. 
The system is submitted to gravity $\mathbf{g}$ (along the $z-$direction) and the bottom wall drives the flow by moving at a given velocity along the $x-$direction.
The top wall is
free to move in the $z-$direction, simply according to the
balance between its weight and the force exerted by the
grains. 
In contrast, it  cannot move along both the $x$ and $y$directions.
Simulations were carried out with $N = 10,000$
slightly polydisperse spheres (uniform number distribution
in the range $0.9\langle d\rangle–1.1\langle d \rangle$) interacting through perfectly
inelastic collisions and Coulomb friction ($\mu = \mu_{pp} = 0.5$). As mentioned above, the 
coefficient of restitution is expected to have nearly no influence on dense
granular flows due to the presence of enduring contacts~\cite{Rajchenbach2000}. Consequently we have chosen perfectly inelastic grains to
maximize dissipation and thus optimize computation time.
Interactions of particles with the flat walls were also
perfectly inelastic and frictional (with a coefficient of
friction $\mu_{pw}$). 
Note that the grain size distribution is narrower than that used for gravity-driven flows. Yet, the results presented here are insensitive to this parameter as long as long range order (obtained for purely monosized grains) and segregation (obtained for large size distribution) are prevented.
Similarly to what has been done for gravity-driven flows, the initial kinetic energy is set to a very large value in order to obtain a SFD state without any visible sign of the initial structure of the packing.
We carried out several simulations varying the following parameters: 
(i) the velocity of the bottom wall $V$, (ii) the weight of the upper
wall $M$, and (iii) the particle-wall friction coefficient $\mu_{pw}$. The
first two parameters are made dimensionless respectively by considering a particle Froude number $\tilde{V}=V/\sqrt{gd}$ 
and the ratio between the mass of the top
wall and the total mass of the grains, $\tilde{M}= M/Nm$, where $m$ is
the average particle mass.

\section{Velocity profiles and shear localization}\label{sec:shear}
\subsection{Streamwise velocity}
\begin{figure}%[htbp]
\begin{center}
\resizebox{0.75\hsize}{!}{%
  \includegraphics{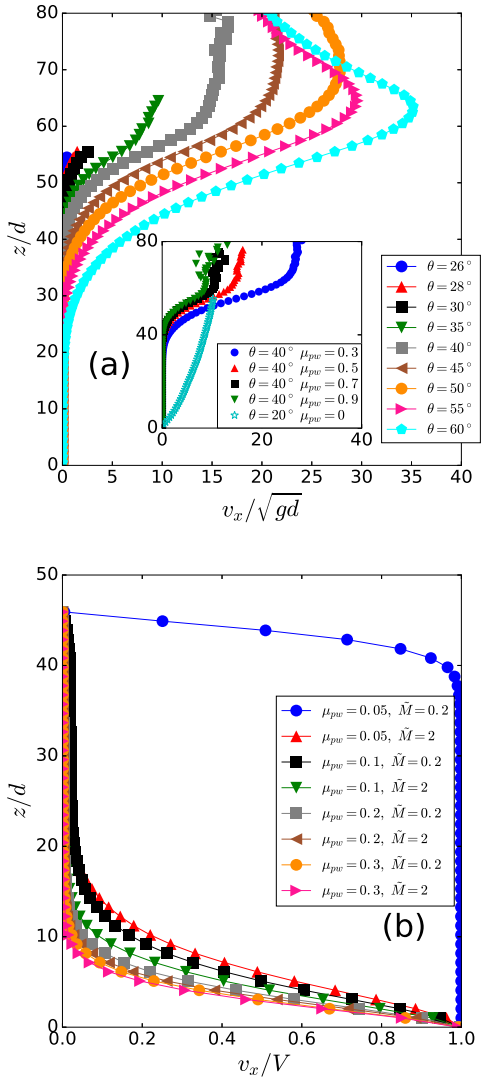}
	}
\caption{
For gravity-driven flows, the vertical profiles of the streamwise velocity show that flow is always localized in the vicinity of the free surface whatever the flow angle (a) and the friction coefficient between grains and sidewalls (inset of (a)). In absence of friction between sidewalls and grains, the systems flows over its whole height. 
In contrast, for shear-driven flows (b), the same profiles demonstrate that in case of large gain-sidewall friction coefficient ($\mu_{pw}$), shear is localized at the bottom of the cell ({i.e.} close to the moving bottom). In contrast, for low values of $\mu_{pw}$ shear is observed in the vicinity of the top of the cell and the rest of the profile is plug-like
}\label{fig:vxvsz}
\end{center}
\end{figure}

\begin{figure}%[htbp]
\begin{center}
\resizebox{0.75\hsize}{!}{%
  \includegraphics{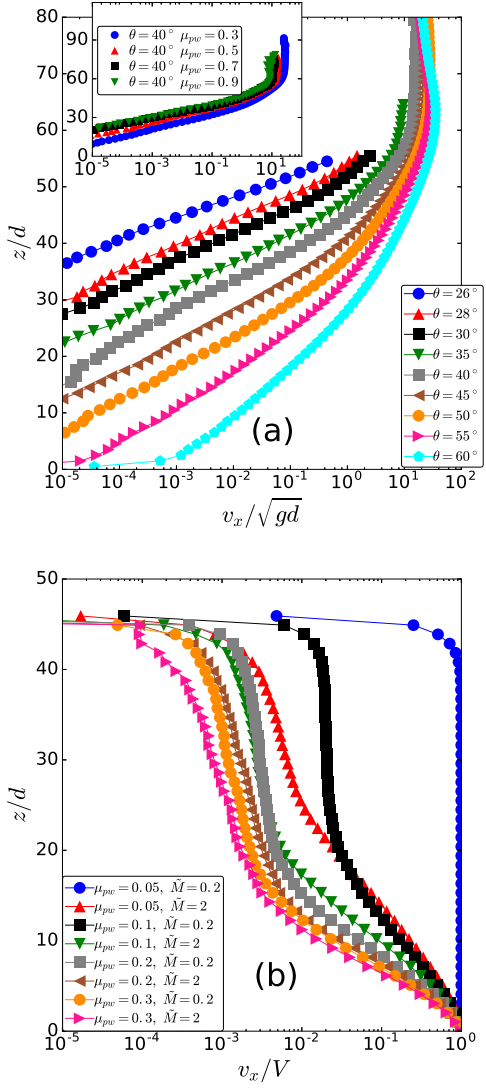}
	}
\caption{
\textcolor{black}{
Both gravity-driven flows ({a}) 
and shear driven flows ({b}) display an exponential velocity profile in the creep zone. Yet, the characteristic lengths are significantly different (a few grains sizes for gravity-driven flows and a few tens of grain sizes for shear-induced flows). 
The effect of the grain-sidewall friction seems to be  weaker for gravity-driven flows (inset of (a) for which the angle of the flow is $\theta=40^\circ$) than for shear-driven flows ({b})}
}\label{fig:logvxvsz}
\end{center}
\end{figure}

\begin{figure}%[htbp]
\begin{center}
\resizebox{0.75\hsize}{!}{%
  \includegraphics{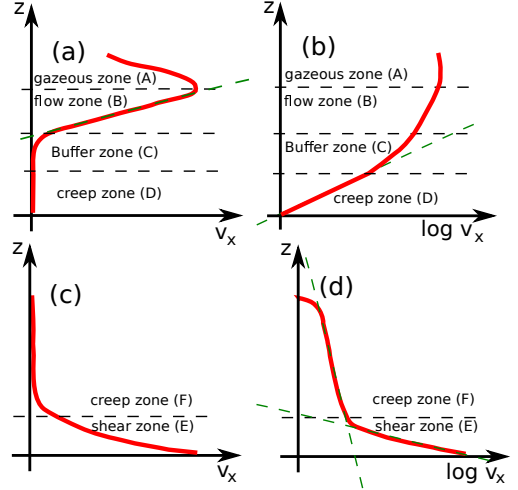}
	}
\caption{Sketch of the velocity profiles for gravity-driven flows ({a}) and ({b}) on a semilog scale. 
Four  zones are reported : zone A or gazeous layer, zone B or flow zone, zone C for buffer zone  and zone D of creep zone.
The same is reported for shear-induced flows ({c}) and ({d}) on a semilog scale
Two zones are reported : zone E and zone F which correspond the shear and creep zones respectively 
}\label{fig:sketch_zones}
\end{center}
\end{figure}

We first focus on the shear localization in the two geometries. Since (i) top boundaries  (free for gravity-driven flows, bumpy wall submitted to a constant pressure for shear-driven flows), (ii) bottom boundaries and (iii) driving forces
are not equivalent in the two geometries, differences are expected.
To address this point we study the vertical profile of the velocity in the main flow direction for the two geometries (see Fig.~\ref{fig:vxvsz}).
Since we focus on steady and fully developed flows, the velocity is averaged over time and along the $x-$direction. Also, unless specified (e.g. for the study of the transverse variations in Sect.~\ref{sec:vtrans}), we average over the $y-$direction.
\textcolor{black}{ 
In agreement with previous studies dealing with gravity-driven flows~\cite{Richard_PRL_2008,Komatsu_PRL_2001,Crassous_JSTAT_2008,deRyck2008b} several zones can be defined from the velocity profiles (see Fig.~\ref{fig:vxvsz}a).
First, a very dilute gazeous zone (zone A) located atop of the flow. 
%It is located below the free surface and above the flow zone defined just below.
%It is characterized by a decrease of $v_x$ with increasing $z$.
Below zone A  is the flow zone (zone B) which is characterized by an almost linear velocity profile. 
To determine its location, we fit linearly the corresponding part of the velocity profile. The upper location for which the velocity profile differs significantly from linearity corresponds to the upper limit of zone B (and thus to the lower limit of zone A). The depth at which the linear fit intercepts the vertical axis corresponds to the lower limit of the flow zone.
Then, below zone B and above the bottom of the system,  we define a buffer zone (zone C) atop a creep zone (zone D)~\cite{Richard_PRL_2008}. In the literature, it has been shown experimentally that the velocity profile in the creep zone exponentially decreases with depth over several decades~\cite{Komatsu_PRL_2001,Crassous_JSTAT_2008}.
}\\
In the chute flow geometry, the flow is always localized at the top whatever the angle $\theta$ (Fig.~\ref{fig:vxvsz}a) and the sidewall friction coefficient (inset of Fig.~\ref{fig:vxvsz}a). This result is expected due to the presence of a free surface. 
Note that, interestingly, flow is localized as long as the sidewall friction coefficient is not zero.
In this case, the system flows over its whole height and the creep zone does not exist anymore (inset of Fig.~\ref{fig:vxvsz}a). Of course, flow angles lower than those obtained with frictional sidewalls are required to obtain  SFD flows.
If we focus on the size of the flow zone (zone B), we can observe that it increases with increasing flow angle and decreasing grain-sidewall friction coefficient. The same is observed for the velocity at a given depth. Discussion on the scaling of the velocity with the flow angle can be found in~\cite{Richard2020}.\\
In the case of shear-driven flows,  the situation is different.
First, it should be pointed out that, for  given $\tilde{M}$ and $\mu_{pw}$, once rescaled by the velocity of the bottom wall, the velocity profiles 
collapse on a single master curve at least for the studied range~\cite{Artoni_JFM_2018}.
For the range of parameters investigated so far
three regimes are observed: (i) for high $\tilde{M}$ and/or high grain-wall
friction coefficient ($\mu_{pw}$), shear is localized at the bottom; (ii) for low $\tilde{M}$ and
low $\mu_{pw}$, shear is localized near the top; and (iii) for
low $\tilde{M}$ and intermediate $\mu_{pw}$, a central plug zone can
form with two shear zones near the bumpy walls. 
It should be pointed out that
in the third case, the shear zone  at the top is very small (a few grain size) and can probably be
interpreted as an apparent slip between the particles and the top wall.
Note also that, in shear zones, velocity profiles are characterized
by an exponential variation whose characteristic length is
mainly a function of $\mu_{pw}$ and $\tilde{M}$~\cite{Artoni_PRL_2015}. 
%If $\tilde{M} \gg 1$, bottom
%localization seems to be more probable with a characteristic
%length decreasing with $\mu_{pw}$; it is reasonable to infer that
%a linear profile should be obtained for sufficiently large $\tilde{M}$
%and sufficiently low $\mu_{pw}$. 
{The possibility for the shear to be localized at different locations within the system  was recently reported for a different
flow configuration~\cite{Moosavi_PRL_2013,Artoni_CompPartMech_2018}}. We have %recently
explained this shear localization by an effective bulk friction heterogeneity~\cite{Artoni_JFM_2018}.
\textcolor{black}{In the remainder of the paper we will mainly focus on the case for which the shear is localized in the vicinity of the bottom, {i.e.} with a shear zone close to the bottom topped by a creep zone. These two zones will be refereed as zone E and zone F respectively.}\\
\textcolor{black}{The question of the boundary between the different zones in gravity-driven flows 
%(\textit{i.e.} between zones B and C)
and the shear and creep zones in shear-driven flows 
({i.e.} zone E and zone F)
is far from being settled. To define a clear boundary between the aforementioned zones, we have reported in Fig.~\ref{fig:logvxvsz} the streamwise velocities on a semilog scale for both geometries (gravity-driven flows in Fig.~\ref{fig:logvxvsz}a and shear-driven flows in Fig.~\ref{fig:logvxvsz}b). In agreement with the literature~\cite{Richard_PRL_2008,Komatsu_PRL_2001,Crassous_JSTAT_2008,Artoni_PRL_2015,Artoni_JFM_2018}, the velocity profile in the creep zones (zones D and F) is exponential for both geometries.
The values of the corresponding characteristic depths are of the order of a few grain sizes in the case of gravity-driven flows and of the order of a few tens of grain sizes for shear-driven flows.\\  
%It should however be pointed out that, for gravity-driven flows, the velocity profile differs significantly from an exponential behaviour in the upper part of the creep zone. 
%This point will be discussed below.\\
%However, as it will be discussed below, in the case of gravity-driven flow and important flow angles, this affirmation has to be handle with care.
The effect of the sidewall-grain friction coefficient on the characteristic length of the velocity profile in the creep zone (zone D) is weak in the case of gravity-driven flows (inset of Fig.~\ref{fig:logvxvsz}a). For shear-driven flows, its effect is more important (Fig.~\ref{fig:logvxvsz}b) and, for a given $\tilde{M}$, the characteristic length of the creeping velocity increases with decreasing $\mu_{pw}$.
These differences suggest that the nature of the creep zone in gravity-driven flows (zone D) is different from that in shear-driven flows (zone F).\\
As mentioned above, in the case of shear-driven flows and shear localization at the bottom of the cell, the velocity profile in the shear zone (zone E) is also exponential. Yet, the characteristic length is significantly smaller than that in the creep zone: a few grains sizes {i.e.} the same order of magnitude than that of the creep zone for gravity-driven flows.
On each velocity profile (still for shear-driven flows with flow localization close to the bottom of the cell) the difference between the two exponentials is clearly visible and the corresponding transition can be used to define the interface between the shear zone and the creep zone.\\
For gravity driven-flows 
the buffer zone (zone C) spans from the
 the depth at which the velocity profile significantly differs from the exponential behaviour to the 
depth at which the flow zone (zone B) %(the zone for which the vertical velocity profile is linear) 
starts~\cite{Richard_PRL_2008}. The size of the buffer zone increases with the angle of the flow. 
%the end of the exponential profiles corresponding to the creep zone is smoother.  
%The depth at which the flow zone (zone B) %(the zone for which the vertical velocity profile is linear) 
%starts might be  larger than 
%the depth at which the velocity profile significantly differs from the exponential behaviour.
%This is especially true for large flow angles.
%Thus, if we defined the flow zone as done before, for large angles, the velocity of the creep zone 
%is only exponential  in its deeper part.\\
%from
% the vicinity of the bottom of the cell to a given height lower than the beginning of the flow zone (zone B).\\
%Yet, the depth at which the velocity profile significantly differs from the exponential behaviour can be used to define the depth of the interface between the creep zone and the flow zone. 
The exact locations of the boundaries between the different zones might appear to be arbitrary~\cite{Richard2020} and  a careful investigation of the grain properties in the vicinity of the interface between zones is probably necessary to quantify them more precisely.}\\
\textcolor{black}{
The different zones defined for gravity-driven flows in the present section ({i.e.} zones A, B, C and D) are sketched in  
Fig.~\ref{fig:sketch_zones}a on a lin-lin scale and in Fig.~\ref{fig:sketch_zones}b  on a semilog scale.
The same is done 
for shear-induced flows and zones E --shear zone-- and F --creep zone-- (Fig.~\ref{fig:sketch_zones}(c)  and 
Fig.~\ref{fig:sketch_zones}(d) on a semilog scale). 
It should be pointed out here an important difference between the two geometries. For shear-driven flows, the upper boundary of the creep zone is defined thanks to the velocity profile on a semilog scale as the point separating two exponential velocity profiles differing by their characteristic length. In contrast, for gravity-driven flows, the profile on a linear-linear scale is mandatory since flow zone is defined as the depth range for which the velocity profile is linear.} 
%This reinforces the aforementioned suggestion according to which the creep zone in gravity-driven flows differ --at last partially-- from that in shear-driven flows.}
\subsection{Transverse velocity profile}\label{sec:vtrans}
\begin{figure}%[htbp]
\begin{center}
\resizebox{0.75\hsize}{!}{%
  \includegraphics{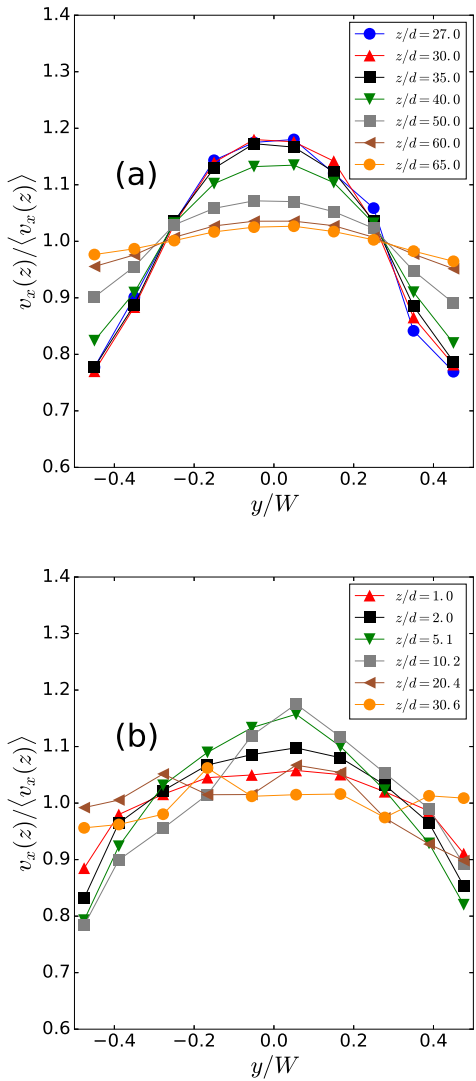}
	}
\caption{The transverse profiles of the relative streamwise velocity ({i.e.} $v_x(z)/\langle v_x(z) \rangle$, where $\langle v_x(z) \rangle$ is the streamwise velocity averaged along the $y-$direction for a given depth $z$) at several depths within a gravity-driven flow ({a}) show that 
the variations of the relative transverse velocity are significant in the creep zone ({i.e.} low $z/d$) and almost negligible at the top of the shear band
({i.e.} high $z/d$). Here the angle of the flow is equal to $\theta=50^\circ$ and the grain-sidewall friction coefficient is set to $0.5$. 
For shear-driven flows (here $\tilde{M} = 2$ and $\mu_{pw}=0.3$, shear localization at the bottom) the opposite is observed ({b}): the relative transverse velocity are the most important in the shear zone ({i.e.} low $z/d$)
}\label{fig:vxvsy}
\end{center}
\end{figure}

In this section we focus on the influence of confinement on the transverse velocity profile. Sidewalls being flat, sliding is expected in their vicinity and, consequently, a transverse plug flow might be observed.  We have reported the transverse velocity profiles for the two geometries (gravity-driven flows in Fig.~\ref{fig:vxvsy}a and shear-driven flows in Fig.~\ref{fig:vxvsy}b) at several depths within the flow. The conditions are ($\mu_{pw}=0.5$; $\theta=50^\circ$) for the gravity-driven flow and ($\mu_{pw}=0.3$; $\tilde{M}=0.2$)  for the shear-driven flow. 
Note that using $\mu_{pw}=0.3$ instead of $\mu_{pw}=0.5$ in former geometry gives similar results.
It should be pointed out that the quantity reported is the relative transverse velocity ($v_x(z) / \langle v_x(z) \rangle$, where $\langle v_x(z) \rangle$ is the streamwise velocity averaged along the $y-$direction for a given depth $z$), consequently we are focusing on transverse relative variations and not the absolute ones. Note that
for shear-driven flows, the set of parameters used leads to a localization of shear at the bottom of the simulation cell.\\
For the gravity-driven flows, we observe that the variations of the relative transverse velocity are significant in the creep zone  (up to $25\%$). They decrease when approaching the free surface  and the transverse profile tends towards a plug flow. It should be pointed out that this situation corresponds to dilute flows (volume fraction lower that $0.3$) that cannot be achieved for shear-driven flows due to the presence of the bumpy wall which applies a constant confining pressure on the top of the granular system. \\
The relative variations of the transverse velocity profiles seem to be similar in the case of shear-driven flows. Yet, in contrast to gravity-driven flows,
largest variations are found for the shear  zone ({i.e.} $z/d=1,\ 2,\ 5\mbox{ and }10$), those of
the creep zone being negligible. This suggests 
that the nature of the creep zones is different in the two geometries and, consequently, is strongly influenced by the boundaries.
More precisely, the important relative variations of the velocity observed in the creep zone of gravity-driven flows and in the shear zone of shear-driven flows suggest that these two zones share common properties.\\
%This point, which will be confirmed in next section, has huge implications for the derivation of boundary conditions for creep zones since the exact geometry of the system should be taken into account.

\section{Granular temperature}\label{sec:T}
Granular temperature is a measure of the velocity fluctuations of the grains. It is a key parameter in many theories aiming to capture granular flows behaviour~(\cite{JenkinsBerzi2010,Zhang_PRL_2017} among many other). It is defined as $T=\left(T_{xx}+T_{yy}+T_{zz}\right)/3$ with $T_{ij}=\langle u_i u_j \rangle - \langle u_i \rangle \langle u_j \rangle$ where, $u_i$ is the component along the $i-$direction of the grain velocity and $\langle \ldots \rangle$ stands for average. Similarly to what have been done for the velocity we averaged our data over time, along the $x-$direction and, unless specified, along the $y-$direction.
\subsection{Vertical temperature profiles}\label{sec:Tz}
\begin{figure}%[htbp]
\begin{center}
\resizebox{0.75\hsize}{!}{%
  \includegraphics{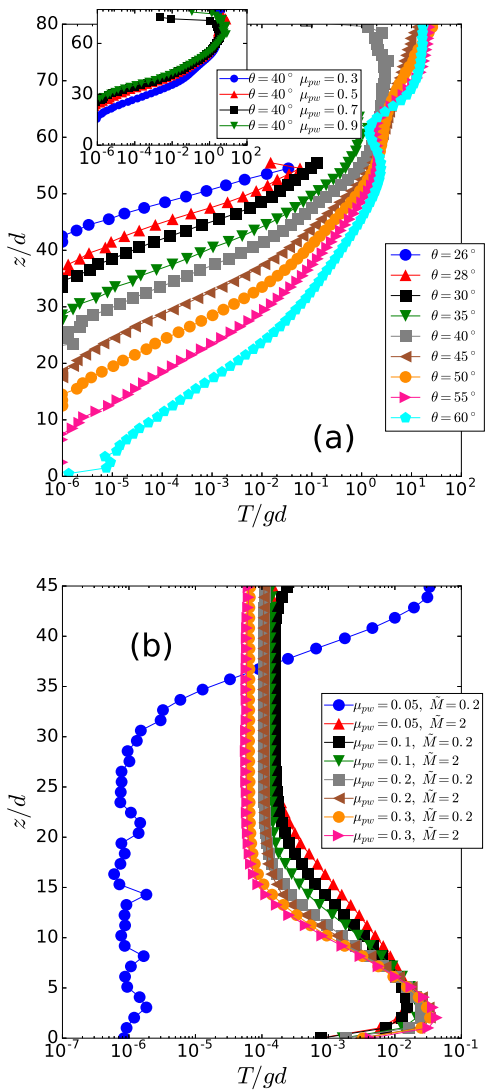}
	}
\caption{
In the gravity-driven flows ({a}) the temperature profiles continuously increases from the  bottom of the system to the end of the flow zone. The  creep zone is dissipative and this dissipative character slightly depends on the friction coefficient between grains and sidewalls (inset of ({a}) for which the angle of the flow is $\theta=40^\circ$). For shear-driven flows (b), if shear is localized close to the bottom, a maximum of temperature is observed in its vicinity (see text). In contrast, when shear is localized at the top of the cell, the maximum of temperature is localized at the top
}\label{fig:Tvsz}
\end{center}
\end{figure}

Figure~\ref{fig:Tvsz} depicts the vertical profiles of the granular temperature for gravity-driven flows (Fig.~\ref{fig:Tvsz}a) and shear-driven flows (Fig.~\ref{fig:Tvsz}({b})).
For the gravity-driven flows, the temperature profiles continuously increase from the  bottom of the system  to the end of the flow zone. 
This variation  is somewhat expected since the   creep zone is dissipative and thus the temperature increases with the distance from this zone. Consequently, the grains are more and more agitated from the bottom of the flow zone to its top, the temperature increases when approaching to the free surface. As mentioned above, in the vicinity of the free surface, very dilute and potentially ballistic flows  can be observed. Their study is out of the scope of this paper.
The temperature profile in the buffer and creep zones is exponential but the characteristic length is larger for the former. Consequently the boundary between these two zones appears clearly on the temperature profile.  
\textcolor{black}{
%The temperature profile can be divided into two parts using the following procedure.
%Starting from the bottom, the profile is 
%exponential but, at a given depth which is still in the creep zone, the profile remains exponential but with a different characteristic length. 
This is different from what we have observed in the streamwise velocity profile for which the corresponding transition was  smoother.
%In the case of gravity-driven flows, we can thus define  two sub-zones: the upper creep zone (zone C1) and the lower creep zone (zone C2) as sketched in figure~\ref{fig:sketch_zones_bis}. The zone C2 corresponds to 
%the part of the zone C for which the velocity profile is exponential.
%\begin{figure}%[htbp]
%\begin{center}
%\resizebox{0.75\hsize}{!}{%
  %\includegraphics{sketch_zones_bis.eps}
	%}
%\caption{Sketch of the temperature profiles for gravity-driven flows on a semilog scale. 
%Three zones are reported : zone A or gazeous layer, zone B or flow zone  and zone C of creep zone.
%The creep zone can de divided into two sub-zones: the lower creep zone (zone C2) and the upper creep zone (C1). In the two zones the temperature profile is exponential but with different characteristic length. 
%}\label{fig:sketch_zones_bis}
%\end{center}
%\end{figure}
Interestingly the values of the characteristic lengths in the creep zones of gravity-driven flows (zone D) is similar to that measured in the flow zones of shear-driven flows (zone E) ({i.e.} a few grains sizes).  
The fact that the characteristic length of the exponential velocity profile of zone E (shear zone of a shear-induced flow) is similar to that of zone  D (creep zone of a gravity-induced flow) strongly suggests that both zones are equivalent. Consequently, what we call the creep zone for shear-driven flows (zone F) does not exist in gravity-driven flows. Similarly the flow zone and, obviously, the gazeous layer observed in gravity-induced 
flows have no counterparts in the shear-induced flows.}\\  
For shear-driven flows, when the shear is localized at the bottom of the cell,
the moving bumpy bottom is a dissipative boundary probably because the grains making up the wall cannot move on relatively to the other.
However, the shear induced by the wall acts as a ``heat source'' in its vicinity. For this reason, granular temperature first increases with $z$, then reaches a maximum for a depth corresponding to a few grain layers above the bottom wall. Far from the latter wall, 
the granular temperature profile reaches a constant value. 
%the moving bumpy bottom is a temperature source and it``heats'' the system. The maximum of the  temperature is then reached  for a depth corresponding to a few grain layers above the bottom then it decreases and reaches a constant value. 
When the shear is localized at the top of the cell, the temperature is constant (and very small) in the creep zone which behaves thus like a dissipative base above which the flow occurs. Consequently, in the shear region, the temperature gradually increases until the top of the cell.\\
The vertical profiles of the temperature show that the properties of the creep zone depend on the geometry. For shear-driven flows, the temperature of the creep zone is constant whereas for gravity-driven flows it continuously decreases with depth. This confirms the results reported in Sect.~\ref{sec:vtrans}.  A possible explanation of these differences is the following: due to the presence of a confining top wall (and thus the absence of a free surface), which is allowed to move vertically,
 shear-driven flows may exhibit non-negligible solid-like fluctuations~\cite{Artoni_JFM_2018} even far from the location of the  top wall. This points out the necessity of non-local modelling of granular flows for which boundary conditions influence the system over long distances. In contrast, for gravity-driven flows, such solid-like fluctuations, if they exist, are located deep in the creep zone and are thus very weak.\\
\textcolor{black}{Interestingly, we can observe a correspondence between the velocity and the temperature vertical profiles. Each zone identified in the velocity profile ({i.e.} shear, creep, plug\ldots and even gazeous zones) can also be easily identified in the temperature profiles. As an example, for shear-induced flows with shear localization close to bottom, the creep zone is defined by an exponential velocity profile between the bumpy bottom and a given depth. Besides, it is defined by a constant temperature between the top wall and the same given depth. Similarly, the shear zone is defined by an exponential velocity profile, with a significantly lower characteristic length with respect to that of the creep zone, and   a non-constant temperature for the same depths.} \\

\subsection{Transverse temperature}\label{sec:Ttrans}
\begin{figure}%[htbp]
\begin{center}
\resizebox{0.75\hsize}{!}{%
  \includegraphics{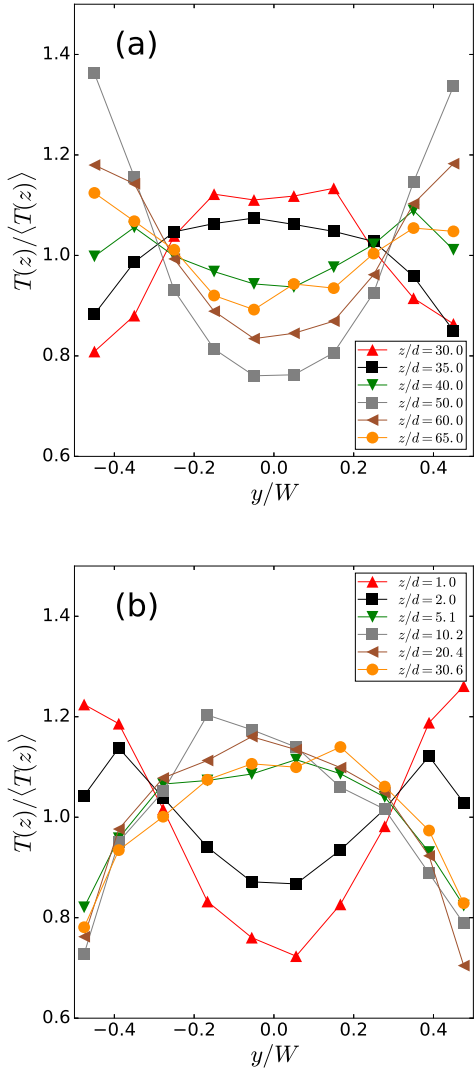}
	}
\caption{
The transverse profiles of the relative streamwise velocity ({i.e.} $T(z)/\langle T(z) \rangle$, where $\langle T(z) \rangle$ is the temperature averaged along the $y-$direction for a given depth $z$) at several depths within a gravity-driven flow show that, 
depending on the depth $z$,
the sidewalls behave like a source or a sink of granular temperature. This is true for both gravity-driven ({a}) 
-- here the
angle of the 
flow is equal to $\theta=50^\circ$ and the grain-sidewall friction
coefficient is set to $0.5$ -- and shear-driven ({b}) flows --here $\tilde{M} = 2$
and $\mu_{pw} = 0.3$, shear localization at the bottom}\label{fig:Tvsy}
\end{center}
\end{figure}

Similarly to what has been done for the streamwise velocity (see Sect.~\ref{sec:vtrans}), we have reported in Fig.~\ref{fig:Tvsy} the transverse profiles of the relative temperature for different depths and for the two geometries  (gravity-driven flow in Fig.~\ref{fig:Tvsy}a and shear-driven flow in Fig.~\ref{fig:Tvsy}b).\\
The transverse profiles of the  granular temperature demonstrate once again the crucial effect of sidewalls on the flow properties.
In the lowest part of the shear zone (for shear-driven flows) and in the flow zone (for gravity-driven flows) this quantity is greatest at the sidewalls and lowest in the centre of the simulation cell.
It should be pointed out that, for shear-driven flows, what we call lowest of the shear zone is probably strongly influenced by the bumpy bottom. It corresponds to the part of the vertical profile of the temperature for which the temperature increases (see Fig.~\ref{fig:Tvsz}). 
In contrast, for both systems 
in the creep zone, the granular temperature gradually rises from its minimal value
at the sidewalls to a maximum value at the centre of the cell.  
The consequences of these results are important.
Depending on the vertical position,  sidewalls
can be either a granular heat source (in the shear zone for shear-driven flows and in the flow zone for gravity-driven flows) or a sink (in the creep
zone). 
This demonstrate the complexity of  stipulating a
sidewall boundary condition on the granular temperature for theories aiming to capture
the properties of granular flows involving both creep and shear/flow zones. 
Yet, this point is crucial since it has been recently shown that the temperature can be used to describe non-local effects~\cite{Zhang_PRL_2017}.
It is worth noting that despite the difference pointed out in Sect.~\ref{sec:Tz} the relative transverse variations of the temperature are similar in the two studied geometries.

\subsection{Scaling with the shear rate}
In the literature, it is common to try to link the granular temperature with the shear rate
$\dot\gamma(z)=\partial \langle v_x(z) \rangle / \partial z$~\cite{Losert_PRL_2000,Mueth_PRE_2003,Orpe_JFM_2007,Zhang_EPJE_2019,Artoni_PRL_2015} to better understand the relation between velocity fluctuations and the  rheology of the system. In the framework of kinetic theory, the granular temperature indeed appears in the expression of the effective viscosity thus on that of the stress. It has been reported a power-law relation between the two quantities {i.e.} $T\propto \dot\gamma^\kappa$ with the power $\kappa$ ranging between approximately $1$ in case of slow and dense  flows and approximately $2$ for fast and dilute flows. The work of Orpe and Khakhar~\cite{Orpe_JFM_2007} is particularly clear and explicit on that point: they show that, in  the case of surface flows in a confined rotating drum, the exponent $\kappa$ increases with the rotation speed of the drum from $\kappa=1$ to $\kappa=2$.
\begin{figure}%[htbp]
\begin{center}
\resizebox{0.75\hsize}{!}{%
  \includegraphics{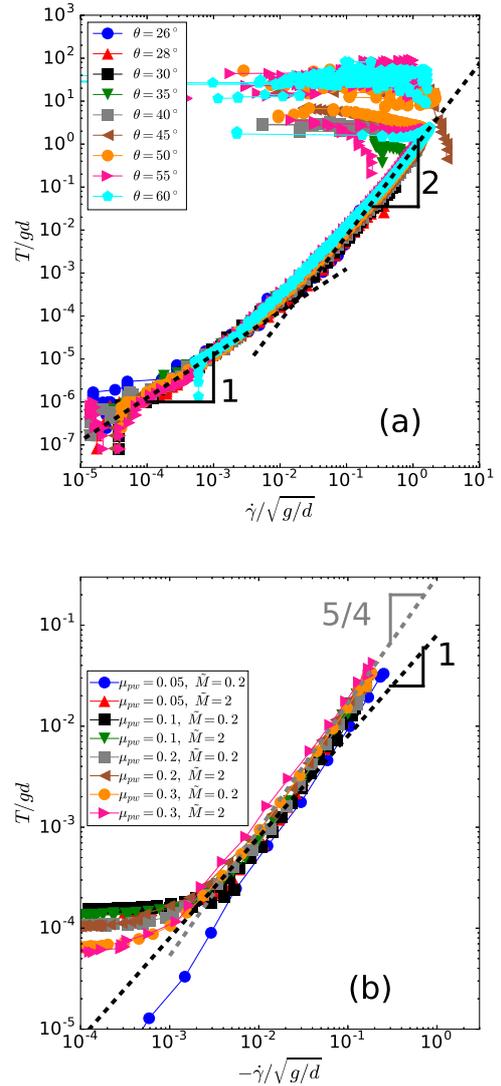}
	}
\caption{For gravity driven flows, the granular temperature, $T(z)$ is linked to the shear rate $\dot\gamma(z)$ through a power law $T(z)\propto \dot\gamma(z) ^\kappa$ where $\kappa$ varies from $1$ (dense flow) to $2$ (dilute flow) ({a}). The latter relation is also valid for shear-driven flows (b) but the maximum value of $\kappa$ obtained for the fastest part of the flow is $\kappa\approx 5/4 ({b})$ 
}\label{fig:Tvsgam}
\end{center}
\end{figure}
For the gravity driven flows (Fig.~\ref{fig:Tvsgam}a), we recover the results obtained by Orpe and Khakhar: we have $T\propto \dot\gamma$ in the creep zone ({i.e.} for low speed) and $T \propto \dot\gamma^2$ at the top of the flow zone ({i.e.} for important speed). 
\textcolor{black}{In other words,  we can write $T\propto \dot\gamma^\kappa$ with $\kappa$ varying between $1$ and $2$.
Yet, $\kappa\approx 2$ for the whole flow zone.
The aforementioned scalings have been obtained with shear-rate and temperature averaged over time and along $x-$ and $y-$directions. Note that we have checked that they still hold if the data are not averaged along the $y-$direction but measured at the sidewalls. 
}\\
The relation between the temperature and the shear rate is also valid for granular flows obtained in the shear cell (Fig.~\ref{fig:Tvsgam}({b})). However, since the latter flows are slower than those driven by gravity, the power $\kappa$ is indeed close to $1$ for the slowest part of the shear zone but cannot reach the value of $2$: for the fastest part of the shear zone we have measured $\kappa\approx 5/4$. \textcolor{black}{This confirms that the latter zone does not correspond to the flow zone (zone B) in gravity-driven flows but is closer to the creep zone (zone D).}
For low shear rates ({i.e.} in the creep zone F) $T/{gd}$ is approximately constant consequence of the plateau of constant granular temperature observed in Sect.~\ref{sec:Tz}. This highlights again the importance of non-locality in creep flows.

%%%%%%%%%%%%%%%%%%%%%%%%%%%%%%%%%%%%%%%%%%%%%%%%%%%%%%%%%%%%%%
\section{Sidewall friction}\label{sec:friction}
In preceding sections we have shown that the presence of frictional and flat sidewalls has a strong influence on the behaviour of granular flows. Yet, we focused only on the kinematic properties. Below we will report sidewall friction measurements, discuss the spatial evolution of this quantity within the system and the relation with the sliding velocity at sidewalls.
\subsection{Friction weakening}\label{sec:weak}
To understand confined granular flows, a key observable is the effective sidewall friction coefficient. It is defined as $\mu_\tau \equiv ||\vec{\tau}_w|| / || \sigma_{yy}^w||$, which we compute as the magnitude ratio of 
the surface force $\vec{\tau}_w \equiv \sigma_{yx}^w \vec{e_x} +\sigma_{yz}^w \vec{e_z}$ and 
normal stress $\sigma_{yy}^w$ on sidewalls, 
\vec{e_x} and \vec{e_z} being unit vectors along the $x-$ and $z-$directions, respectively.\\
\begin{figure}%[htbp]
\begin{center}
\resizebox{0.75\hsize}{!}{%
  \includegraphics{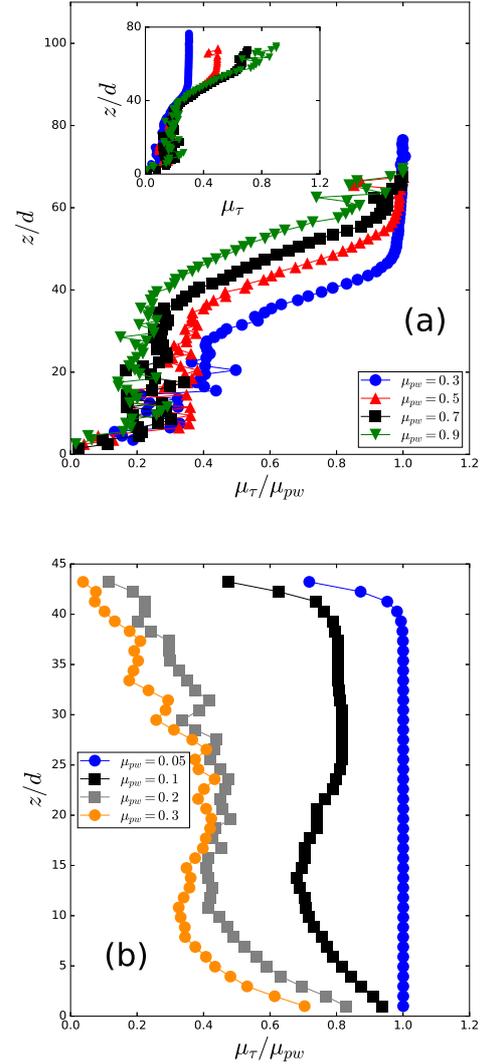}
	}
\caption{The effective friction coefficient on sidewalls, $\mu_{\tau}$, is strongly influenced by confinement. For gravity driven flows with an angle of the flow $\theta=40^\circ$ ({a}), in the vicinity of the free surface, it is close to the grain-sidewall friction coefficient, $\mu_{pw}$, especially for low values of the latter. Then,  it weakens and, in the creep zone, reaches a constant value $\mu_\tau \approx 0.17$
which is independent of the grain-sidewall friction coefficient (inset of ({a})). 
For shear-driven flows (b), when the shear is localized at the top of the cell, $\mu_\tau\approx \mu_{pw}$ in the plug-flow region and then decrease in the shear band. When shear is localized at the bottom of the cell,  $\mu_{\tau}$ approaches $\mu_{pw}$ at the end of the shear zone ({i.e.} close to the bottom) and then decreases. In both types of flows a significant increase can be observed in the middle of the creep zone
 }\label{fig:fric_weak}
\end{center}
\end{figure}
For gravity-driven flows the effective friction coefficient is close to microscopic particle - wall friction coefficient in the flow zone (Fig.~\ref{fig:fric_weak}a and its inset).
This is especially true for low particle-wall friction coefficients,  the Coulomb threshold $F_t = \mu_{pw} F_n$ being more easily achieved
in this case.
 Deeper in the flow, the effective friction weakens and tends towards a constant value ($\approx 0.17$) in the creep zone. This value is independent of the grain-sidewall friction coefficient (see the inset of Fig.~\ref{fig:fric_weak}a). 
%This can be easily understood since 
In the creep zone, the Coulomb threshold is far from being reached  and the corresponding ratios 
\textit{tangential force to normal force} remains below $\mu_{pw}$.
For shear-driven flows
the evolution of the effective friction is strongly related to velocity profiles. If shear is localized at the top of the cell, far from the shear band, grains move as a plug in
the $x-$direction with velocity $V$ and  all the grains in contact with sidewalls slip. Thus, the effective coefficient of friction in the plug zone is equal
to $\mu_{pw}$. In the shear zone, stick-slip events may emerge and a friction lower than $\mu_{pw}$ is observed.\\ 
As mentioned above, for both geometries, the effective sidewall friction weakens in the creep zone (see Fig.~\ref{fig:fric_weak}) where
%It should be pointed out that another important quantity is the orientation of the sidewall friction (\textit{i.e.} the angle between $\sigma_{yx}$ and $\sigma_{yz}$)
the friction also has a component along the  vertical direction (see~\cite{Richard_PRL_2008,Artoni_PRL_2015}).
Note also that this peculiar behaviour has been recently observed experimentally~\cite{Artoni_JFM_2018} for shear-driven flows.
This friction weakening  can be
explained by the fact that, reasonably, stick-slip events
become more and more probable when we approach the
creep zone.
Then, significant slip events become less frequent deeper in
the creep zone, thereby increasing the time during which grains describe a random oscillatory
motion with zero mean displacement~\cite{Richard_PRL_2008}. The latter behaviour contributes
negligibly to the mean resultant wall friction force.
This result demonstrates 
that boundary conditions for dense granular flows must support the possibility of
%the necessity to use a
non-constant effective friction coefficient between sidewalls and the system.
{Note also that Yang et al.~\cite{Yang_granularmatter_2016} derived a model 
 describing how  the weakening of $\mu_\tau$ can be related to the
ratio between rotation-induced velocity and sliding velocity.}

\subsection{Sliding at sidewalls}
\begin{figure}%[htbp]
\begin{center}
\resizebox{0.75\hsize}{!}{%
  \includegraphics{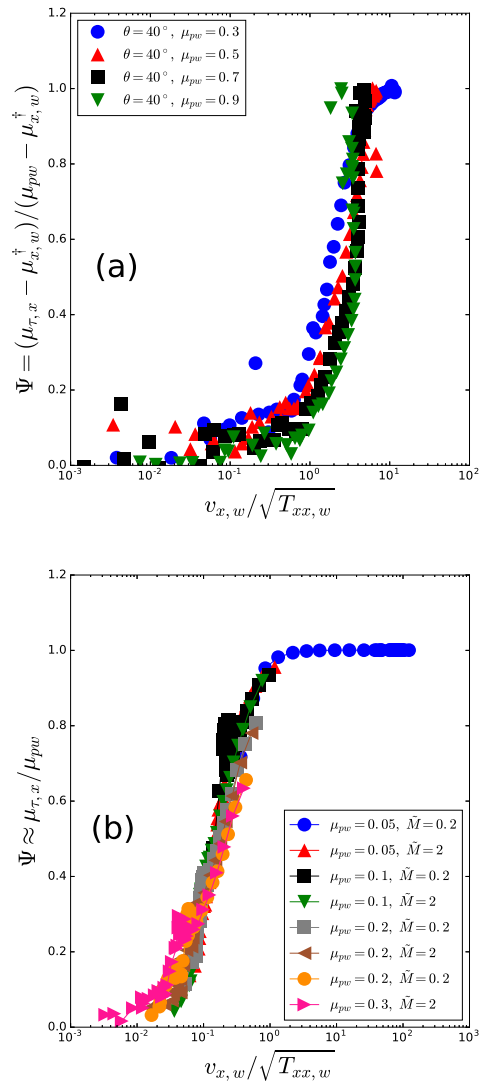}
	}
\caption{
The rescaled effective wall friction coefficient in the main flow direction, $\Psi$,
is linked to the dimensionless slip parameter $v_{x,w}/\sqrt{T_{xx,w}}$ for gravity-driven flows ({a}) and shear-induced flows ({b}). 
For gravity-driven flows, since effective friction does not tend towards zero when approaching the bottom of the flow, the rescaled 
effective wall friction coefficient is %\mu_{0,x}
$\Psi=(\mu_{\tau,x} - \mu_{\tau,x}^\dagger)/(\mu_{pw} - \mu_{\tau,x}^\dagger)$ where $\mu_{\tau,x}^\dagger$ is the value of  $\mu_{\tau,x}$
when $\uslide$ tends towards zero.
For shear-induced flows the rescaled 
effective wall friction coefficient is approximated by $\mu_{\tau,x}/\mu_{pw}$
}\label{fig:slip}
\end{center}
\end{figure}

Theories aiming to describe granular flows require boundary conditions. For that purpose several authors have used the ratio $v_x/\sqrt{T_{xx}}$ in kinetic theories~\cite{johnson_jackson_1987,Richman1988} or extended kinetic theories~\cite{JenkinsBerzi2010,Jenkins_granularmatter_2012}. 
Also, since it has been recently shown that the granular fluidity ({i.e.} the ratio of the pressure to the effective viscosity) scales with the square root of the granular temperature
 a connection also exists with the nonlocal
theories recently developed~\cite{Kamrin_PRL_2012,Bouzid_PRL_2013,Kamrin_SoftMatter_2015,Zhang_PRL_2017}. Moreover, for these theories, the question of the boundary conditions to be used is still open.  
To test these approaches we have reported (Fig.~\ref{fig:slip}) for the two configurations the evolution of the rescaled effective sidewall friction in the $x-$direction ({i.e.} 
$\Psi=(\mu_{\tau,x}-\mu_{\tau,x}^\dagger)/(\mu_{pw}-\mu_{\tau,x}^\dagger)$ where $\mu_{\tau,x}=\left|\left|\sigma_{yx}^w\right|\right|/\left|\left| \sigma_{yy}^w \right|\right|$ 
and $\mu_{\tau,x}^\dagger=\mu_{\tau,x}$ for $\uslide$ tends towards zero) 
 {with the following ratio : $\uslide$, for which the subscript $w$ stands for  quantities at sidewalls.} 
Note that for shear-induced flows, $\mu_{\tau,x}^\dagger\approx 0$ and $\Psi$  can be approximated by $\mu_{\tau,x}/\mu_{pw}$.  
In both systems (gravity-driven and shear-driven flows)
the scaling performs globally well on several orders of
magnitude. 
The two master curves obtained are similar: $\mu_\tau$ increases with $\uslide$ and a plateau is potentially reached when friction is fully mobilized ($\mu_\tau = \mu_{pw}$) for high values of $\uslide$. Yet the variations of $\mu_{\tau,w}$ at low values of $\uslide$  show that the  shape of the master curve is clearly geometry-dependent. 
Sidewall friction coefficient indeed tends towards zero at low $\uslide$ for shear-driven flows (when shear is localized at the bottom) whereas for gravity-driven flows, it reaches a plateau whose value is not zero. A striking point should be pointed out. The reported curves have an S-shape, indicating that a strong increase of $\mu_{\tau,x}/\mu_{pw}$ is observed for a relatively small variation of $v_{x,w}/\sqrt{T_{xx,w}}$. This indicates an important  correlation between  $v_{x,w}$ and $\sqrt{T_{xx,w}}$ during the aforementioned increase.
%due to the presence of gravity (see section~\ref{sec:weak}).
It should also be pointed out that some deviations from the
master curve are observed. In particular the sidewall friction coefficient seems to have a visible effect. Also, for shear-driven flows, slight variations are observed due to the nature of the regime ({i.e.} shear localized at the bottom, at the top or both at the bottom and the top of the cell).

This strongly suggests that velocity fluctuations play an important role for the boundary conditions and should be a key parameter in theoretical description of granular flows.  
This point is consistent with the recent observation that ``fluidity'' and granular temperature are strongly linked~\cite{Zhang_PRL_2017}.

%As shown in~\cite{Artoni_PRL_2015} a possible form of the equation linking $\mu_x/\mu_{pw}$ with $\uslide$ is $\mu_{w}/\mu_{pw} = \left(\uslide \right)^B / \left( A + \left(\uslide \right)^B\right)$.

\section{Conclusion}
We have studied the properties of two types of confined flows. The first case is a flow on a bumpy bottom driven by gravity  and confined between two flat but frictional sidewalls. The second case corresponds to a flow confined between not only two flat and frictional sidewalls but also between a bumpy top and a bumpy bottom. It is  driven by shear induced by the bottom wall moving at constant velocity. 
In addition to the driving of the flow, these two situations differ by their boundary conditions at their top ({i.e.} respectively a free surface condition and a constant pressure condition) and bottom (respectively zero velocity and constant velocity).  
As a consequence the former geometry leads to much looser flows than the latter. 
\textcolor{black}{We have identified in each type of flow different zones: gazeous, flow, buffer and creep zones for gravity-driven flows and shear and creep zones for shear-driven flows with shear localization at the bottom.
Our results suggest that the shear zone of a shear-induced flow does not correspond to the flow zone of a gravity-driven flow but to its creep zone.}\\
We have shown that in both conditions, the lateral confinement is of great importance. 
In the case of gravity-driven flows, flow is localized at the free surface whatever the grain-sidewall friction coefficient as long as it has a finite value.
%The height of the flow zone increases when the value of  grain-sidewall friction coefficient decreases.
%In the case of chute flows the friction of sidewalls induces a shear localization close to the free surface. 
The case of shear-induced flow is  more complex.  For low grain-sidewall friction coefficient, shear is localized in the vicinity of the bumpy top. In contrast for important values of the latter coefficient, it is localized close to the moving bumpy bottom which drives the flow. Also, the relative transverse variations of the velocity are different. For the gravity-driven flows, in the vicinity of  the top of the flow, 
the relative transverse variations of the velocity is weak,
%the transverse velocity profile is close to a plug 
whereas they are close to $25\%$ in the creep zone. For shear-induced flows and a shear localization at the bottom of the simulation cell, the opposite is observed (low relative variations in the creep zone, important ones in the shear zone).
Also, the vertical profile of the granular temperature in the shear cell shows a plateau which could be induced by the bumpy top wall. This suggests the presence of long range effects in granular flows and demonstrates the necessity to introduce non-local effects in their theoretical description.\\
The two systems share common properties. 
%First the velocity  are sheared in the transverse direction in the creep zone and this shear decreases when approaching the flow zone for which the transverse velocity profile is close to a plug. Second, 
First, in both cases, the transverse profiles of the granular temperature show that sidewall could be either a granular heat source or a sink. This demonstrates the difficulty to write a simple boundary condition at sidewalls for granular temperature. Second, in both cases the scaling of the granular temperature with  shear is similar confirming that the latter relation can be used to quantify the rheology of the system. 
Finally, in both cases, sidewall effective friction (i) weakens in the creep zone, consequence of the intermittent motion of the grains~\cite{Richard_PRL_2008,Artoni_PRL_2015}  and (ii) seems to be linked to a slip parameter defined as $\uslide$.\\
Our results demonstrate the importance of studying boundaries in granular flows and shed light on the complexity of such a study. 
They also suggest that a full three-dimensional
rheological description of a granular flow is  required.
Yet, the studied
geometries, by the complexity of the flows they produce and by the importance of the
boundaries, are well adapted for testing granular rheologies numerically and studying
boundary conditions. In particular, the presence of boundaries highlights the importance
of non-local effects on flow behaviour %(Kamrin & Koval 2012; Bouzid et al. 2013,
%2015; Kamrin & Henann 2015; Nott 2017; Zhang & Kamrin 2017); 
\cite{Kamrin_PRL_2012,Bouzid_PRL_2013,Kamrin_SoftMatter_2015,Nott_EPJWebConf_2017}
our systems are
thus relevant to test the theories taking into account the latter effects.

\begin{acknowledgement}
We thank Ph. Boltenhagen for fruitful discussion on granular chute flows. The numerical simulations were carried out at the CCIPL
(Centre de Calcul Intensif des Pays de la Loire).
\end{acknowledgement}

\section*{Compliance with ethical standards}
\textbf{Conflict of interest} The authors declare that they have no conflict of
interest.
\bibliographystyle{unsrt}
\bibliography{dem8}

\end{document}